\documentclass[10pt,letterpaper]{article}
\usepackage{opex3}
\usepackage{amsmath}
\usepackage{SIunits}
\usepackage{cite}

\begin{document}

\title{Optical refrigeration with coupled quantum wells}

\author{Rapha\"{e}l~S.~Daveau,$^{1,*}$ Petru~Tighineanu,$^1$ Peter~Lodahl,$^1$ and S\o ren~Stobbe$^1$}

\address{$^1$Niels Bohr Institute, University of Copenhagen, Blegdamsvej 17, DK-2100 Copenhagen, Denmark}

\email{$^*$raphael.daveau@nbi.ku.dk} 



\begin{abstract}
Refrigeration of a solid-state system with light has potential applications for cooling small-scale electronics and photonics. We show theoretically that two coupled semiconductor quantum wells are efficient cooling media for optical refrigeration because they support long-lived indirect electron-hole pairs. Thermal excitation of these pairs to distinct higher-energy states with faster radiative recombination allows an efficient escape channel to remove thermal energy from the system. This allows reaching much higher cooling efficiencies than with single quantum wells. From band-diagram calculations along with an experimentally realistic level scheme we calculate the cooling efficiency and cooling yield of different devices with coupled quantum wells embedded in a suspended nanomembrane. The dimension and composition of the quantum wells allow optimizing either of these quantities, which cannot, however, be maximized simultaneously. Quantum-well structures with electrical control allow tunability of carrier lifetimes and energy levels so that the cooling efficiency can be optimized over time as the thermal population decreases due to the cooling.
\end{abstract}



\section{Introduction}
Refrigeration is the cooling of a system by a heat-exchange mechanism and can be achieved through interaction with laser light \cite{Pringsheim1929}. Temperature and heat are closely related to vibrations in solids (phonons). The field of phononics is striving to exploit and tailor the interaction between phonons and matter, exemplified by the recently proposed thermal transistors able to control and modulate heat flux on small length scales \cite{ben2014}. Phonons are, however, a nuisance for many applications, in particular for quantum sciences where they lead to decoherence mechanisms \cite{Lodahl2013}. Laser cooling is nowadays a robust way to cool the mechanical motion of atoms and ions in dilute gases and is at the heart of ultracold atomic-physics experiments \cite{cohen1998}. The effects of radiation pressure are exploited to cool individual mechanical modes of microscopic objects towards the quantum-mechanical ground state in quantum optomechanics \cite{Aspelmeyer2014}. In these experiments, one single mode is cooled but the effective temperature of the system is generally not decreased. In contrast to the research directions outlined above, the goal of optical refrigeration is to cool all mechanical modes of an entire structure, which results in temperature and entropy reduction. This could lead to a compact solid-state refrigerator able to reach cryogenic temperatures with light \cite{Sheik2007} without cryogenic fluids or moving parts.

Optical refrigeration in solids relies on up-conversion luminescence (also referred to as anti-Stokes luminescence, which must not be confused with nonlinear optical effects, such as parametric up-conversion), which blue-shifts the emitted light with respect to the light entering the system in the first place. This process removes thermal energy and therefore leads to refrigeration \cite{Sheik2007}. The cooling induced per up-converted photon is of the order of the thermal energy $k_BT\sim 26$ meV at room temperature. On the other hand, if the absorbed laser photon is lost non-radiatively, the heating induced is of the order of the band gap energy for a semiconductor, typically about 1 eV. Therefore, net cooling of a solid can only take place if the total radiative efficiency of the emission $\eta_T > 1-k_BT/\hbar\omega_L$ \cite{Epstein2009}, where $\omega_L$ is the laser photon frequency. In practice, high-purity materials with near-unity radiative efficiency are required to fulfill this condition. While the inherent inefficiency of optical refrigeration is bound to a level of a few per cent ($\sim k_BT/\hbar\omega_L$) it could be used for cooling small-scale structures, e.g, in quantum technology.

Since the theoretical prediction by Pringsheim almost a century ago \cite{Pringsheim1929}, optical refrigeration has been primarily observed in rare-earth doped glasses or crystals \cite{Epstein1995,Mungan1997,Thiede2005} and is now able to attain cryogenic temperatures \cite{Seletskiy2010,Melgaard2013}. More recently, optical refrigeration of a semiconductor was achieved in CdSe nanostructures \cite{zhang2013}. There are several advantages of using semiconductors for refrigeration: they can, in principle, be cooled to lower temperatures and they can remove heat faster because of the fast radiative recombinations inherent to semiconductors \cite{Epstein2009}.

Although the theory behind optical semiconductor refrigeration relying on broad acoustic-phonon sidebands to remove heat \cite{sheik2004} has been widely addressed, the experimental results are still lacking. The first successful cooling experiment of a semiconductor was achieved in a nanostructure and used discrete optical-phonon replicas \cite{zhang2013}. As opposed to these two approaches, our proposal exploits a quantum heterostructure with quantized energy states, which can be approximated by a 3-level scheme as shown in Fig.~\ref{fig:figure1}(a). Optical pumping of carriers into a long-lived excited state located a few $k_BT$ in energy below a much faster recombination channel \cite{Khurgin2006} can lead to an efficient heat extraction as the carriers recombine through the higher-energy state after absorbing thermal energy. Indirect and direct electron-hole pairs confined in coupled quantum wells (CQWs) in an electric field fulfill this criterion of long and short lifetime, respectively.

The lifetime of the indirect excitons in CQWs can be enhanced by at least two orders of magnitude by means of applied electric fields \cite{Butov1999,golub1990}. This indicates that their quantum efficiency can exceed 99\%, because any non-radiative effects would set a lower bound to how much the lifetime can be reduced. This renders CQWs promising for optical refrigeration. Local carrier cooling effects have been observed with indirect excitons in CQWs structures \cite{high2009} but the cooling of the entire crystal lattice was neither intended nor achieved. As opposed to previous theoretical works on optical refrigeration with single quantum wells (QWs), we propose here for the first time using CQWs to cool an entire nanostructure.

We find that by optimizing the thickness and alloy composition of the CQWs embedded in a suspended nanomembrane, a temperature drop of 80 K from room temperature is predicted with 1 W of laser power, corresponding to a cooling efficiency per photon $\eta_\mathrm{p}$ of around 3\%. Although this cooling efficiency is comparable to that of other semiconductor platforms or rare-earth doped glasses \cite{Seletskiy2010}, the tunability of carrier lifetimes and energy levels with an applied bias allow to optimize the cooling efficiency over time as the thermal population $k_BT$ decreases with cooling.

\section{Optical refrigeration with indirect electron-hole pairs}
Coupled quantum wells in a constant electric field along the growth direction support essentially two different types of electron-hole pairs (EHPs): direct and indirect \cite{Butov1999,Snoke2000}. Electrons and holes located in the two different QWs form indirect EHPs with small spatial overlap resulting in a long lifetime. Electrons and holes located in the same QW form direct EHPs with a strong spatial overlap and correspondingly shorter lifetime. A similar configuration of indirect and direct EHPs can be achieved using type-II QWs \cite{Epstein2009}. Electron-hole pairs confined in CQWs are versatile since an external electric field can be used to tune the energy levels and the spatial overlap between EHPs, which are essential parameters in the optical refrigeration scheme as shown below.

\begin{figure}[h!]
  \centering
  \includegraphics[width=\textwidth]{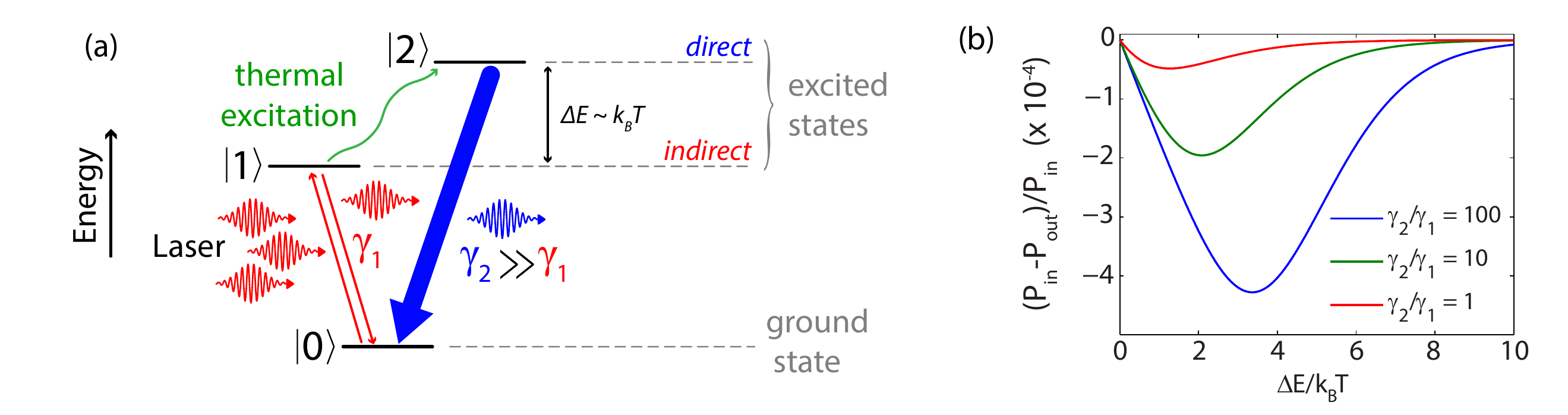}
  \caption{(a) Principle of optical refrigeration of a semiconductor with CQWs. Photo-excited indirect EHPs are thermally excited to a higher-energy state with a faster recombination rate ($\gamma_i$ indicate the recombination rate of the different states). This process removes thermal energy as photons leave the system with a higher energy than the absorbed photons. (b) Normalized net power deposited on the system $P_\mathrm{net} = P_\mathrm{in} - P_\mathrm{out}$ as a function of energy spacing between the direct and indirect states for different lifetime ratios.  Larger $\Delta E$ (cooling efficiency) can be achieved for EHPs with small spatial overlap in the ground state.}
  \label{fig:figure1}
\end{figure}

We model the 3-level scheme from Fig.~\ref{fig:figure1}(a) as follows. Carriers are optically pumped from the ground state $|0\rangle$ to the indirect EHP states $|1\rangle$ with the rate $\gamma_\mathrm{p}$. Acoustic phonons mediate the interaction between the excited states with the rates $\gamma_\mathrm{ph}$ from $|2\rangle\to |1\rangle$ and $ \gamma_\mathrm{ph}^\star = \gamma_\mathrm{ph}\exp(-\Delta E/k_B T)$ from $|1\rangle\to |2\rangle$, where $\Delta E = E_2 - E_1$ is the energy difference between $|2\rangle$ and $|1\rangle$ and $k_B$ is the Boltzmann constant. The two excited states $|1\rangle$ and $|2\rangle$ recombine radiatively with the rates $ \gamma_1$ and $ \gamma_2 $, respectively. Under continuous excitation with power $P_\mathrm{laser}$, the pumping rate is given by $ \gamma_\mathrm{p} = f(\alpha) P_\mathrm{laser}/E_1  $, where $ f(\alpha) $ represents the fraction of incoming photons absorbed by the system. For the sake of simplicity we first assume $ f(\alpha) = 1 $, but its value will be calculated accurately later on for realistic structures. The two populations $ N_1 $ (indirect) and $ N_2 $ (direct) satisfy the following system of equations,

\begin{equation}
\frac{d}{dt}
\begin{pmatrix}
N_{1} \\
N_{2}
\end{pmatrix}
=
\begin{pmatrix}
\gamma_\mathrm{p} \\
0
\end{pmatrix}
+
\begin{pmatrix}
- \gamma_{1} - \gamma_\mathrm{ph}\mathrm{e}^{-\beta\Delta E} & \gamma_\mathrm{ph}\\
\gamma_\mathrm{ph}\mathrm{e}^{-\beta\Delta E} & - \gamma_{2} - \gamma_\mathrm{ph}
\end{pmatrix}
\begin{pmatrix}
N_{1} \\
N_{2}
\end{pmatrix}
=
\begin{pmatrix}
0 \\
0
\end{pmatrix}
\end{equation}
where $ \beta = 1/k_BT$ and the last equality holds for steady-state condition. The solution reads
\begin{equation}
 N_2 =  \frac{\gamma_\mathrm{ph}\gamma_\mathrm{p}}{\gamma_\mathrm{ph}\gamma_2 + \gamma_1(\gamma_2+\gamma_\mathrm{ph})\exp(\beta\Delta E)} \qquad \mathrm{and} \qquad N_1 = \frac{\gamma_\mathrm{p} - \gamma_2 N_2}{\gamma_1}.
\end{equation}
The acoustic phonon scattering rate is generally much faster than the radiative recombination rates for bulk semiconductors above 10 K \cite{oberhauser1993,sheik2004}, which is the relevant temperature for optical refrigeration, so we assume $ \gamma_\mathrm{ph} \gg \gamma_1,\gamma_2 $. Thus, the quantities above become
\begin{equation}
N_2\simeq\frac{\gamma_\mathrm{p}}{\gamma_2 + \gamma_1\exp(\beta\Delta E)} \qquad \mathrm{and} \qquad N_1 \simeq \frac{\gamma_\mathrm{p}}{\gamma_\mathrm{1}}\left(1 - \frac{\gamma_2}{\gamma_2 + \gamma_1\exp(\beta\Delta E)}\right).
\label{eq:N1N2}
\end{equation}
The net power deposited is defined as $ P_\mathrm{net} = P_\mathrm{in} - P_\mathrm{out} = f(\alpha) P_\mathrm{laser} - \left( \gamma_{2}N_2 E_2 + \gamma_{1}N_1 E_1 \right) $ so that $P_\mathrm{net} < 0 $ is the condition for achieving optical refrigeration. Figure~\ref{fig:figure1}(b) plots the normalized $ P_\mathrm{net}$ as a function of $\Delta E$ for different values of the lifetime ratio $\gamma_2/\gamma_1$. The optimal energy spacing between the two excited states, corresponding to the position at the minimum of each curve, is of the order of a few $k_BT$ and is increasing with the ratio $\gamma_2/\gamma_1$. Additionally, Eq.~(\ref{eq:N1N2}) shows that, in the limit where $\gamma_1\ll\gamma_2$, $N_2/N_1\approx \exp(-\beta\Delta E)$, indicating that the photon flux emitted from the higher-energy state $ \gamma_2N_2 $ will likely overcome that of the lower-energy state $ \gamma_1N_1 $. Therefore, having long-lived indirect EHPs satisfying $\gamma_2/\gamma_1 \gg 1$ is an important requirement for achieving an efficient cooling and optimizing the cooling efficiency, defined as $\Delta E/\hbar\omega_L$. For a typical value $\Delta E \sim 4k_BT \simeq 100$ meV, the cooling efficiency of a 1 eV photon is 10\% at room temperature. This is a significant improvement as compared to 4.8\% in CdSe nanostructures \cite{zhang2013} or 3.5\% in type-II QWs \cite{Epstein2009}. As shown below, however, the best cooling performance is not obtained by maximizing the cooling efficiency because it is necessarily associated with a low absorption. In the next section, we explore the optical refrigeration of a realistic microstructure with embedded CQWs where lifetimes and energies of indirect and direct states can be accurately tuned.

\section{Cooling efficiency and cooling yield of coupled quantum wells}

\begin{figure}[t]
  \centering
  \includegraphics[width=\textwidth]{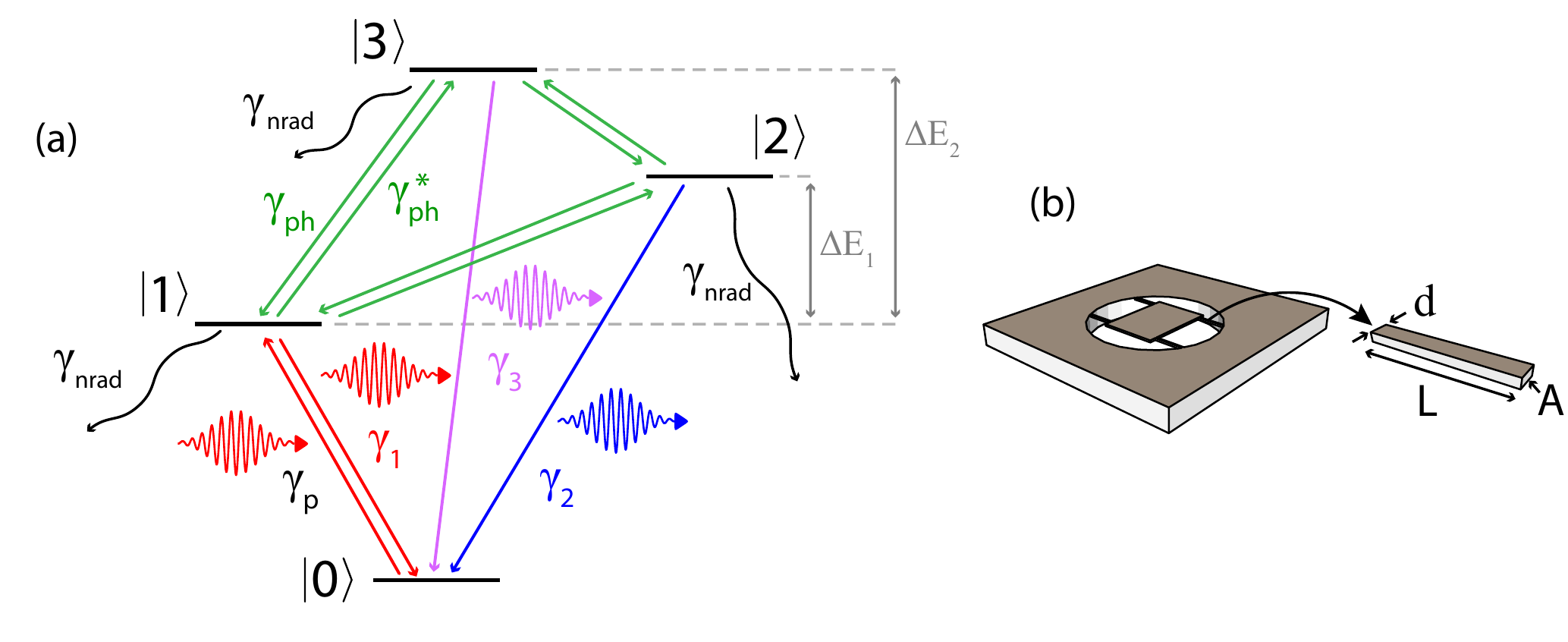}
  \caption{(a) Four-level scheme characterizing the first few EHP transitions of CQWs. The long-lived indirect EHP state $\lvert 1\rangle$ is pumped optically and the two higher-energy direct states $\lvert 2\rangle$ and $\lvert 3\rangle$ are thermally populated. (b) Design of a realistic structure for achieving optical refrigeration experimentally. The dimensions of the tethers suspending the membrane allows to estimate the thermal resistance and the achievable cooling yield.}
  \label{fig:figure2}
\end{figure}

The phenomenological quantities introduced in the rate equations above are computed microscopically in order to account realistically for the creation, thermal redistribution and recombination of the carriers. The band diagram of CQWs is modelled and the eigenstates are calculated by solving the single-particle effective-mass equation (solved with a tunnelling resonance technique \cite{Tighineanu2011} with band parameters taken from Ref.\ \cite{Vurgaftman2001}). Using this model, transition energies, radiative and non-radiative recombination rates, and absorption coefficients can be calculated. For the type of CQWs considered in this work, one indirect and several direct states separated by a few $k_BT$ appear in the band diagram simulations. We therefore consider a more general 4-level system as shown in Fig.~\ref{fig:figure2}(a). The non-radiative recombination rate $\gamma_\mathrm{nrad}$ is assumed to be constant for all states and equal to the tunnelling rate of carriers out of CQWs. However, since these rates are very small as compared to radiative ones, the results do not depend significantly on them. The system of rate equations for the three populations $ N_1 $ (indirect) and $ N_2 $, $N_3$ (direct) reads
\begin{equation}
\frac{d}{dt}
\begin{pmatrix}
N_{1} \\
N_{2} \\
N_3
\end{pmatrix}
=
\begin{pmatrix}
\gamma_\mathrm{p} \\
0 \\
0
\end{pmatrix}
+
M_4
\begin{pmatrix}
N_{1} \\
N_{2} \\
N_3
\end{pmatrix}
=
\begin{pmatrix}
0 \\
0 \\
0
\end{pmatrix}
\end{equation}
where the last equality only holds for steady-state condition and
\begin{equation*}
M_4=
\begin{pmatrix}
- \gamma_{1} - \gamma_\mathrm{nrad} - \gamma_\mathrm{ph}(\xi_1+\xi_2) & \gamma_\mathrm{ph} & \gamma_\mathrm{ph}\\
 \gamma_\mathrm{ph}\xi_1 & - \gamma_{2} - \gamma_\mathrm{nrad} - \gamma_\mathrm{ph}(1 + \xi_2/\xi_1) & \gamma_\mathrm{ph}\\
 \gamma_\mathrm{ph}\xi_2 & \gamma_\mathrm{ph}\xi_2/\xi_1 & - \gamma_{3} - \gamma_\mathrm{nrad} - 2\gamma_\mathrm{ph}
\end{pmatrix}
\end{equation*}
where $\Delta E_1 = E_2 -E_1$ and $\Delta E_2 = E_3 -E_1$, and we have introduced the notation $ \xi_i = \exp(-\beta\Delta E_i) $.

\begin{figure}[t]
  \centering
  \includegraphics[width=\textwidth]{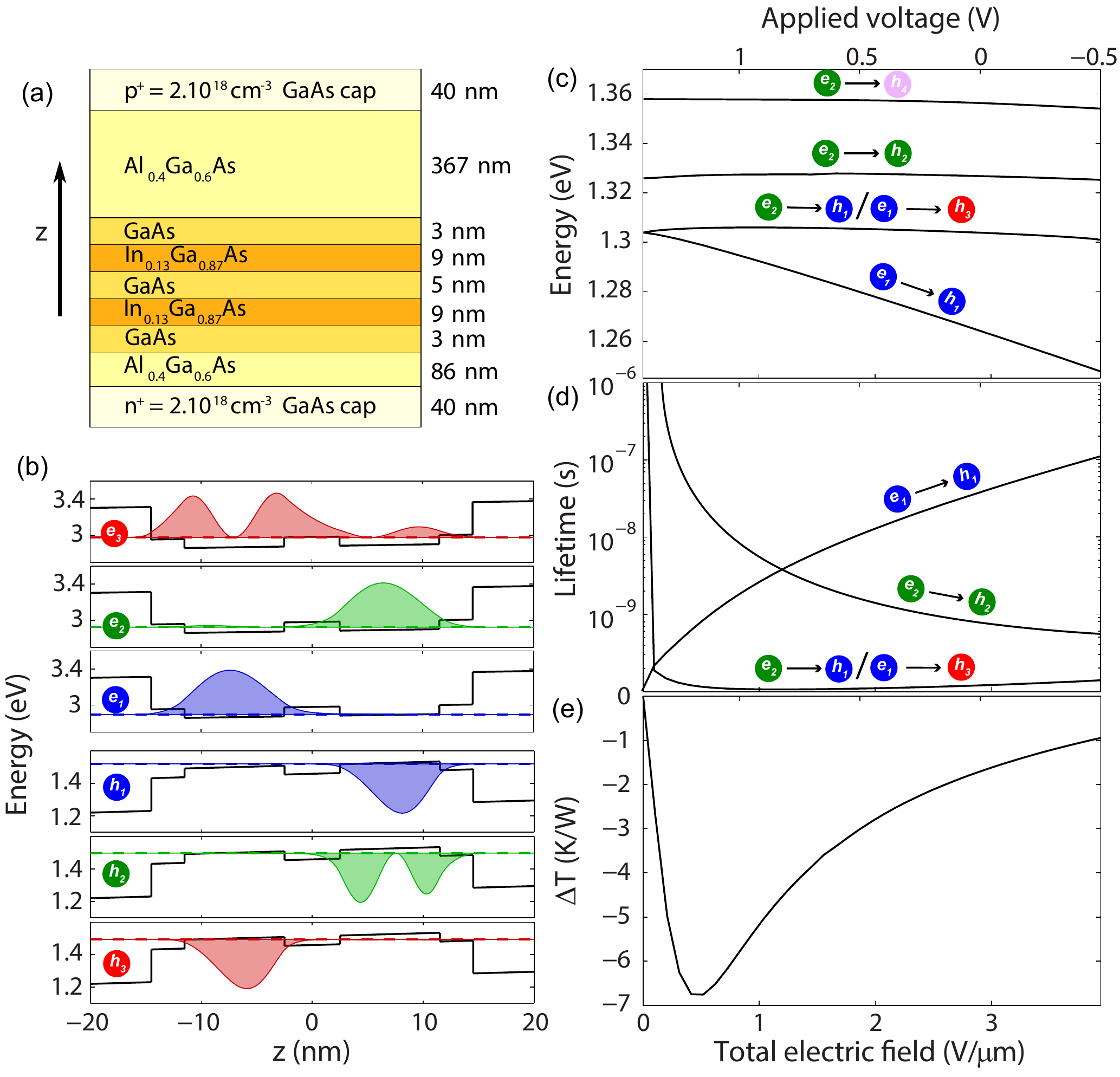}
  \caption{Membrane design with high cooling efficiency (design 1). (a) Layer structure of the suspended membrane. (b) Computed band diagram for $V = 0$ V along with the first three electron and hole probability densities. Note that the wavefunctions of $e_1$ and $h_1$ overlap very little in space. The radiative decay rate is directly proportional to the spatial overlap between the electron and hole wavefunctions. This transition has therefore a longer lifetime as compared to the direct transitions $e_2\to h_1$ and $e_1\to h_3$, which has a larger overlap. (c) Energy levels for the first few transitions. (d) Corresponding lifetimes. (e) Cooling yield as a function of total electric field.}
  \label{fig:figure3}
\end{figure}

One of the limiting factors for the cooling of a realistic device is the inflowing heat from the surrounding environment. We model this by considering a suspended membrane containing the CQWs as depicted in Fig.~\ref{fig:figure2}(b), so as to thermally insulate the cooled area from its environment. In practice, such a suspended membrane can be fabricated by standard processes involving wet and dry chemical etching \cite{midolo2015}. The substrate below the membrane may even be completely removed using selective etching so as to leave only the membrane itself in the optical path \cite{Liu2011}. Another effect with potentially detrimental consequences is parasitic background absorption, which can convert an incoming or outgoing photon directly into heat. This is particularly critical for bulk materials because the photon-escape probability is low and thus the re-absorption probability is high. Our proposed device is designed to counteract parasitic absorption by two measures: 1) The submicron thickness of the membrane ensures a short optical path with essentially single-pass operation. We note that in the work by Zhang et al. \cite{zhang2013}, a model assuming a unity escape probability was found to faithfully model their experiment using a similar geometry. 2) As opposed to optical refrigeration based on phonon-sideband excitation, the optical transitions of CQWs occur at energies well below the absorption threshold of the surrounding materials, which significantly reduces the parasitic absorption. For these reasons, we assume negligible re-absorption and parasitic absorption in the further analysis.

The steady-state temperature drop per watt of laser power, i.e. the cooling yield $\Delta T$ is estimated through $\Delta T = R_\mathrm{th} P_\mathrm{net}$ where $R_\mathrm{th} = \rho L/A = 5.81$ cm$\cdot$K/W is the thermal resistance of the four tethers of length $L = 20 \;\micro\meter$ and cross-sectional area $A = 0.5 \times 1 \;\micro\meter^2$ suspending the membrane. The values for the thermal resistivity $\rho$ of the different alloys were taken from \cite{adachi1983,afromowitz1973}. For the 4-level system of Fig.~\ref{fig:figure2}(a), $ P_\mathrm{net} = f(\alpha) P_\mathrm{laser} - \left( \gamma_{3}N_3 E_3 + \gamma_{2}N_2 E_2 + \gamma_{1}N_1 E_1 \right) $, where the absorption coefficient $ f(\alpha) $ of the $\lvert 0\rangle \to \lvert 1\rangle$ transition is calculated from the transition matrix element. We compute $\Delta T$ self-consistently, by taking into account that the thermal population $\propto \exp(-\Delta E_i/k_BT)$ changes as the temperature decreases. Indeed, $\Delta T$ would be overestimated if only calculated in one iteration because the thermal population is larger at room temperature than at the target temperature. Therefore, we calculate $\Delta T$ iteratively until its value has converged.

\begin{figure}[t]
  \centering
  \includegraphics[width=\textwidth]{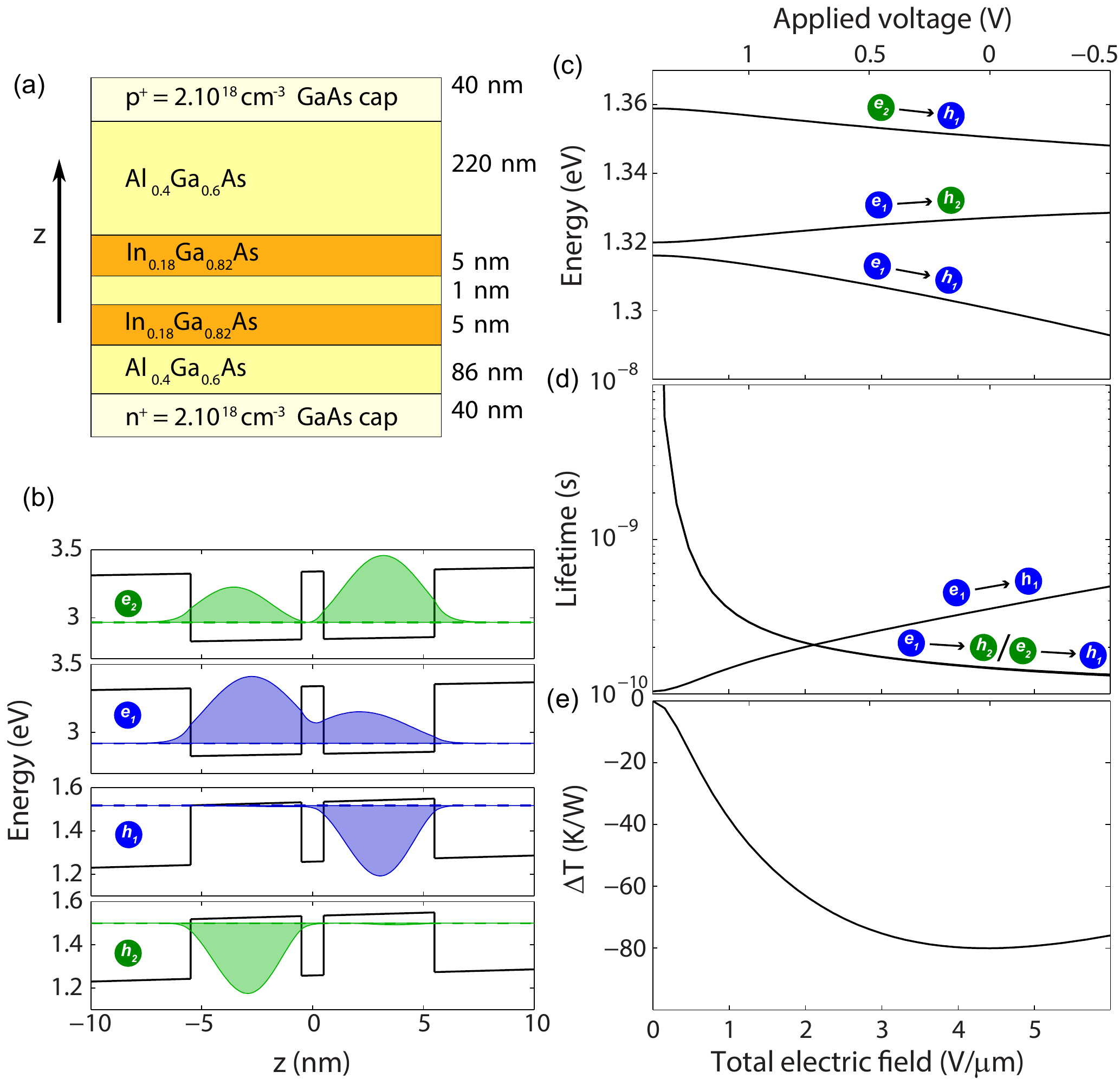}
  \caption{Membrane design with optimized cooling yield (design 2). (a) Layer structure of the suspended membrane. (b) Computed band diagram for $V = 0$ V along with the first two electron- and hole-eigenstate probability densities. (c) Energy levels for the first few transitions. (d) Corresponding lifetimes. (e) Cooling yield as a function of total electric field.}
  \label{fig:figure4}
\end{figure}

We consider two realistic designs of CQWs. The first design is based on a structure that we have explored optically (experiments to be reported elsewhere), and the second design is an improvement of the first in order to optimize the cooling yield. The first design contains two In$_{0.13}$Ga$_{0.87}$As QWs as shown in Fig.~\ref{fig:figure3}(a). The relatively thick potential barrier between the two QWs ensures that electrons and holes in opposite QWs overlap very weakly so that $\gamma_1$ is considerably smaller than $\gamma_2,\gamma_3$. The CQWs are embedded in the intrinsic region of a p-i-n structure so that applying a voltage across the intrinsic region changes the electric field at the position of the CQWs \cite{Fox2001}. Figure~\ref{fig:figure3}(b) shows the electron and hole probability densities in the CQWs for the first few eigenstates. The strength of each transition is proportional to the spatial overlap between the electron and hole wavefunctions. The indirect EHP has the lowest transition energy and is formed by the first eigenstate of the electron $e_1$ and first eigenstate of the hole $h_1$. The next transitions are direct and they are formed by $e_2\to h_1$ and $e_1\to h_3$, which are close in energy, followed at higher energies by $e_2 \to h_2$ and $e_2 \to h_4$. The transition energies and recombination lifetimes are displayed in Fig.~\ref{fig:figure3}(c) and \ref{fig:figure3}(d). The energy of the indirect transition is red-shifting and its lifetime is increasing with electric field due to the quantum-confined Stark effect \cite{Harwit1987}. As a consequence, $ \Delta E_i $ and $\gamma_i/\gamma_1$ can be accurately tuned with voltage in order to achieve the largest cooling yield. For no applied voltage (i.e., only the intrinsic electric field across the CQWs), $\gamma_2/\gamma_1 \sim 500$ and $\Delta E_1 \sim 2k_BT$, which yields a cooling efficiency per photon of $\eta_p \sim 3$\%. Recombinations from higher-energy direct transitions, $e_2 \to h_2$ and $e_2 \to h_4$, lead to $\eta_p \sim 5$\% and 8\%, respectively. This is beyond the experimental values reported for the cooling of CdSe nanostructures \cite{zhang2013}.

Despite the promising fulfilment of the requirement $\gamma_2 \gg \gamma_1$, the reduced overlap of the indirect transition decreases the absorption of pump photons and therefore limits the cooling yield as shown in Fig.~\ref{fig:figure3}(e). The absorption can be increased by reducing the electric field with an applied voltage resulting in an enhanced cooling yield. However, this reduces the cooling efficiency by decreasing $\Delta E_i$ and $\gamma_i/\gamma_1$, i.e., by making the EHPs more direct. There is a trade-off between the benefit resulting from increased absorption and the decrease in the cooling efficiency. Thus there is an optimal configuration in which the cooling yield $\Delta T$ is maximized, which is reached at $V=1.3$ V. Here, the absorption $f(\alpha) $ is about two orders of magnitude larger than at 0 V but the radiative recombination rates $\gamma_1$ and $\gamma_2$ are also comparable, thus the advantageous property $\gamma_2 \gg \gamma_1$ cannot be exploited. In this configuration the achievable cooling is -7 K for 1 W of laser power.

We now investigate the possibility to increase the cooling yield with an optimized design, simplified to a InGaAs/AlGaAs type of CQWs. A wide range of parameters are varied, such as the QW thickness, barrier thickness and alloy compositions, which all influence energy levels, recombination lifetimes and absorption, until the highest cooling yield is achieved. We restrict the optimization parameters such that the eigenstate of the InGaAs CQWs are well below the absorption threshold of the GaAs barriers to avoid unwanted absorption elsewhere than in the CQWs, as discussed previously. The resulting design is shown in Fig.~\ref{fig:figure4}(a) while in Fig.~\ref{fig:figure4}(b), the probability densities of the first few electron and hole eigenstates are plotted. The QWs are narrower (5 nm each) so that the indirect-to-direct energy spacings are increased, while the barrier is decreased to 1 nm, resulting in an enhanced intrinsic absorption of the CQWs ($ f(\alpha) \sim 10^{-2} $). The achievable cooling yield versus voltage is shown in Fig.~\ref{fig:figure4}(e) and reaches -80 K at 0 V. Figure~\ref{fig:figure4}(c) indicates that at 0 V, where $\Delta T$ is largest, $\Delta E_1 = 33$ meV and $\Delta E_2 = 53$ meV, which gives a cooling efficiency in the range of 2.3 to 3.8\%.

\section{Discussion}

\begin{figure}[b]
  \centering
  \includegraphics[width=\textwidth]{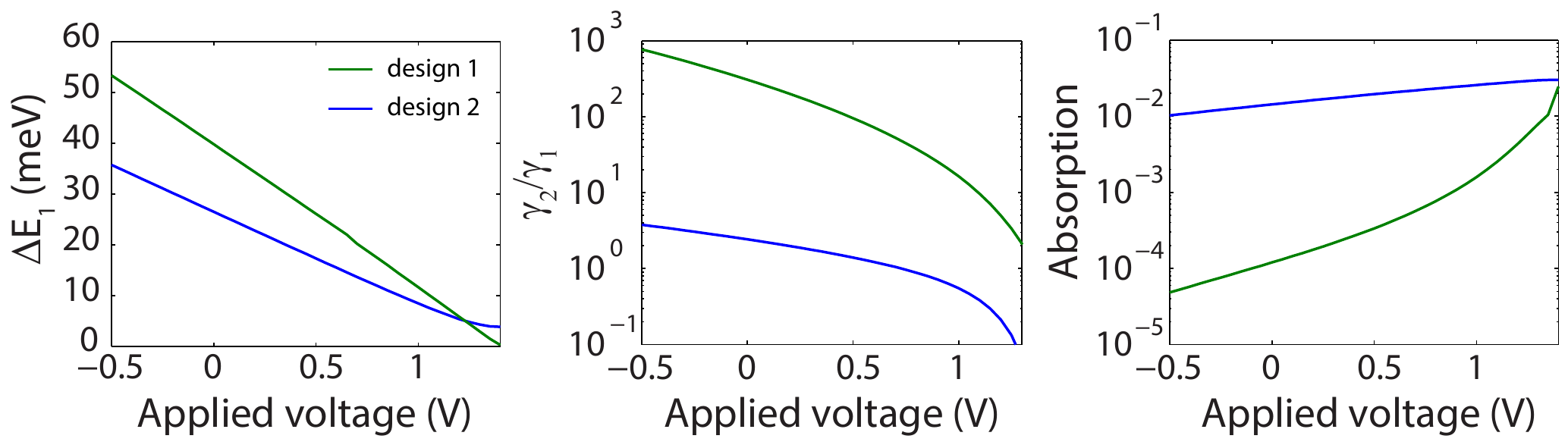}
  \caption{Comparison of the parameters of interest for optical refrigeration for the two designs. (a) Energy spacing between the indirect and first direct state $\Delta E_1$. (b) Lifetime ratio between the indirect and first excited state $\gamma_2/\gamma_1$. (c) Absorption of the suspended membrane. There is a trade-off between, on the one hand, increasing $\gamma_2/\gamma_1$ and $\Delta E_1$, which improve the cooling efficiency and, on the other hand, enhancing the absorption of the membrane, which leads to an increased cooling yield.}
  \label{fig:figure5}
\end{figure}

The configuration allowing the highest cooling yield shows that it is sufficient that $\gamma_1$ is smaller but not much smaller than $ \gamma_2,\gamma_3 $ (see Fig.~\ref{fig:figure4}(d)). This is because a large cooling yield is incompatible with $ \gamma_1 \ll \gamma_2,\gamma_3 $. Figure~\ref{fig:figure5} illustrates this aspect: the first design yields a regime where $\gamma_2/\gamma_1 > 100$ that allows an efficient thermal excitation up to $4k_BT$ above the energy of the indirect state, cf. Fig.~\ref{fig:figure1}(b), but the low absorption of the membrane limits the cooling yield. The second design provides an optimal trade-off between the cooling efficiency and the cooling yield. The latter is the relevant quantity for the experimental realisation of optical refrigeration. This is favoured by the enhanced absorption of the membrane despite the lower cooling efficiency. It means that more photons are involved in the refrigeration process but that each photon removes little thermal energy. In this case, the influence of non-radiative losses or parasitic heating increases \cite{Epstein2009}. Additionally, total internal reflection limits the escape of emitted light from the sample, thus increasing the probability of parasitic absorption. This issue can be prevented by using a solid immersion lens \cite{Gauck1997} on the sample surface or by shaping the density of optical states at the frequency of the re-emitted photons by a second-order Bragg grating \cite{Erdogan1992} or a photonic crystal \cite{Wierer2004}.

\section{Conclusion}

We have proposed using CQWs for optical refrigeration. They exhibit a favorable combination of properties originating from the very different radiative lifetimes for direct and indirect electron-hole pairs. We have shown that CQWs, whose dimensions and composition have been optimized, can be cooled by 80 K with 1 W of optical power from room temperature with an associated cooling efficiency per photon of around 3\%. This cooling efficiency is similar to that of other semiconductor platforms or rare-earth doped glasses. Despite the moderate cooling efficiency, an advantage of CQWs is the ability to tune the lifetime and energy of the different transitions by applying a voltage across the structure. This allows cooling to lower temperatures because the energy levels can be tuned dynamically to maintain optimum refrigeration as the thermal population drops due to the decreasing temperature. Our cooling yield is below the reported value for CdSe nanostructures of 6.3 K/mW in \cite{zhang2013} but higher than that of experiments on rare-earth doped glasses, which use high laser powers \cite{Seletskiy2010}.

Coupled quantum wells offer a great opportunity to tailor the cooling efficiency and the cooling yield as opposed to previous proposals concerned with single QWs. The initial requirement of having $\gamma_2\gg\gamma_1$ in order to achieve efficient optical refrigeration is unfortunately achieved by reducing the recombination rate of the indirect state rather than enhancing that of the direct states. This results in a low absorption and subsequent limited cooling yield. We have investigated several ways of improving the absorption such as using a micro-cavity \cite{Coldren2012} or using side excitation \cite{Epstein2009} to increase the path length of photons in the sample. However, side excitation creates a gradient in carrier concentration, which, due to screening, would lead to a varying electric field. This makes resonant excitation impractical as the spectral position of the indirect EHP state varies at different places in the membrane. Likewise, a cavity favours photon re-absorption, thus setting a more strict requirement on the quantum efficiency. These considerations favour single-pass structures where the light propagates in the same direction as the applied electric field, as considered in the present work. The use of CQWs in a suspended membrane could lead to a successful refrigeration experiment using a high power laser spread on a large area (i.e., with low intensity) because the suspended membrane is expected to yield very little parasitic heating \cite{zhang2013}. The success of the experiment could also be eased by further increasing the thermal resistance of the tethers holding the suspended region to be cooled.

\section*{Acknowledgments}

We would like to thank the Lundbeck Foundation, the Carlsberg Foundation, and the European Research Council (ERC consolidator grant "ALLQUANTUM") for financial support.

\end{document}